\documentclass[11pt,fleqn]{article}
\usepackage{amsfonts}
\usepackage{amsmath,amssymb}
\usepackage{latexsym}

\usepackage{comment}

\usepackage{amsthm}
\usepackage[bookmarks]{hyperref}
\usepackage[capitalize,nameinlink]{cleveref}
\usepackage{caption,subcaption}
\usepackage{pdfpages}
\usepackage{tikz}
\usetikzlibrary{through}

\usepackage{graphicx}
\usepackage{wrapfig}
\usepackage{tikz}
\usetikzlibrary{shapes,backgrounds}
\usetikzlibrary{decorations.pathreplacing}
\usetikzlibrary{arrows,decorations.pathmorphing,backgrounds,fit,positioning,shapes.symbols,chains}
\usetikzlibrary{arrows}
\usepackage{enumitem}
\usepackage{float}
\usepackage{fancybox}
\usepackage{color}
\usepackage{xcolor}

\usepackage{times,tcolorbox}
\newtcbox{\myovalbox}[1]{colback=#1!30,boxrule=0pt,arc=5pt,  boxsep=0pt,left=3pt,right=3pt,top=3pt,bottom=3pt}

\usepackage{tikz}
\usetikzlibrary{intersections,positioning,shapes,calc,arrows,decorations.pathmorphing,decorations.pathreplacing}

\tikzset{iNode/.style={draw=blue, rectangle}}
\tikzset{fNode/.style={draw=green, circle}}
\tikzset{rNode/.style={draw=red, circle}}
\tikzset{nNode/.style={draw, circle}}

\theoremstyle{remark}

\newtheorem*{remark*}{Remark}

\newtheorem*{question*}{Question}

\usepackage{epsfig}

\usepackage{color,soul}
\definecolor{lightblue}{rgb}{.30,.30,1}
\definecolor{lightred}{rgb}{1, .30, 0.30}
\definecolor{lightgreen}{rgb}{.10, 1, 0.10}
\definecolor{darkorange}{rgb}{1.0, 0.55, 0.0}
\definecolor{blue-violet}{rgb}{0.54, 0.17, 0.89}
\soulregister\ref7

\DeclareMathOperator{\KS}{\mathrm{C}\mskip 1mu}
\DeclareMathOperator{\KP}{\mathrm{K}\mskip 1mu}
\DeclareMathOperator{\KM}{\mathrm{KM}\mskip 1mu}
\DeclareMathOperator{\KA}{\mathrm{KA}\mskip 1mu}
\DeclareMathOperator{\KT}{\mathrm{CT}\mskip 1mu}
\DeclareMathOperator{\poly}{\mathrm{poly}\mskip 1mu}
\newcommand{\cnd}{\mskip 2mu | \mskip 2mu}
\newcommand{\cln}{\mskip 2mu{:}\mskip 2mu}

\newcommand{\srp}{{search-to-profile\,}}

\newcommand{\openq}[1]{\begin{enumerate}[resume*] \item #1 \color{black} \end{enumerate}}

\begin{document}

\def\firstcircle{(0,0) circle (1.5cm)}
\def\secondcircle{(0:2cm) circle (1.5cm)}
\def\thirdcircle{(-60:2cm) circle (1.5cm)}

\if01
\centerline{\bf Open Problems Column}
\centerline{\bf Edited by William Gasarch}

\centerline{\bf This Issue's Column!}
This issue's Open Problem Column is by 
FILL IN AUTHORS. 
It is {\it PUT IN TITLE}.
\bigskip

\centerline{\bf Request for Columns!}
I invite any reader who has knowledge of some area to contact me
and arrange to write a column about open problems in that area.
That area can be (1) broad or narrow or anywhere in between,
and (2) really important or really unimportant or anywhere in between.

\bigskip

\fi
\title{27 Open  Problems in Kolmogorov Complexity}

\author{Andrei Romashchenko\thanks{LIRMM, Univ Montpellier, CNRS, Montpellier, France, \url{https://www.lirmm.fr/\~romashchen/}}\\
\small Univ. Montpellier\\[-0.5ex]
\small \texttt{andrei.romashchenko@gmail.com}
\and
Alexander Shen\thanks{ LIRMM, Univ Montpellier, CNRS, Montpellier, France, \url{https://www.lirmm.fr/\~ashen/}}\\
\small Univ. Montpellier\\[-0.5ex] 
\small \texttt{sasha.shen@gmail.com}
\and
{Marius Zimand\/}
\thanks{  Department of Computer and Information Sciences, Towson University,
Baltimore, MD. \url{http://orion.towson.edu/\~mzimand/}; Partially supported by NSF grant CCF 1811729.} \\
\small Towson University \\[-0.5ex]
\small \texttt{mzimand@towson.edu}
}
\date{}
\maketitle

\section{Introduction}

Shannon defined the entropy of a random variable $M$ with $k$ values $m_1,\ldots,m_k$ having probabilities $p_1,\ldots,p_k$ as
\[
H(M) = \sum_i p_i \log \frac{1}{p_i}
\]
This formula can be informally read as follows: the $i$th message $m_i$ brings us $\log(1/p_i)$ ``bits of information'' (whatever this means), and appears with frequency $p_i$, so $H$ is the expected amount of information provided by one random message (one sample of the random variable). Moreover, we can construct an optimal uniquely decodable code that requires about $H$ (at most $H+1$, to be exact) bits per message on average, and it encodes $i$th message by approximately $\log (1/p_i)$ bits, following the natural idea to use short codewords for frequent messages. This fits well the informal reading of the formula given above, and it is tempting to say that the $i$th message ``contains $\log (1/p_i)$ bits of information.'' Shannon himself succumbed to this temptation~\cite[p. 399]{shannon} when he wrote about entropy estimates and considers Basic English and James Joyces's book ``Finnegan's Wake'' as two extreme examples of high and low redundancy in English texts. But, strictly speaking, one can speak only of entropies of random variables, not of their individual values, and ``Finnegan's Wake'' is not a random variable, just a specific string. Can we define the amount of information in individual objects?

The algorithmic information theory (AIT) (developed in the 1960s by Kolmogorov and others) tries to go in this direction. A program $p$ (identified with some canonical encoding of it as a string) for a string $x$ is a program that on empty input prints $x$. AIT defines the amount of information in a finite object (say, a bit string) $x$ as the minimal length of a program for $x$. This quantity depends on the programming language used, but as Solomonoff, Kolmogorov and Chaitin noted, there exist optimal programming languages (sometimes called \emph{universal machines}, though this name can be used for many different notions) that make the complexity function minimal up to $O(1)$ additive terms. One then fixes some optimal programming language and uses it to define the Kolmogorov complexity $\KS(x)$ of a bit string $x$. 

Shannon defined also the \emph{conditional} entropy $H(X\cnd Y)$ of a random variable $X$ relative to some other random variable $Y$. Here we assume that $X$ and $Y$ are defined on the same probability space; Shannon's definition can be equivalently rewritten as $H(X,Y)-H(Y)$. In algorithmic information theory it is natural to define the conditional complexity $\KS(x\cnd y)$ for bit strings $x$ and $y$ as the minimal length of a program that transforms input $y$ to output $x$ (again we have to consider the optimal programming language).\footnote{%
   More formally the conditional complexity can be defined as follows. Let $I$ be an arbitrary algorithm that gets two binary strings $p$ and $x$ and produces (if it terminates) a binary string. This algorithm is intuitively understood as an interpreter of some programming language: $I(p,y)$ is considered as the result of applying program $p$ to input $y$. Then, for a given $I$, we define the function
  \[
  \KS_I(x\cnd y)=\min\{|p|\colon U(p,y)=x\}
  \]
  Solomonoff and Kolmogorov observed that there exists an optimal $I$, i.e., an algorithm $I$ such that for every other algorithm $I'$ the inequality
  \[
  \KS_I(x\cnd y)\le \KS_{I'}(x\cnd y)+c 
  \]
  holds for some $c$ and for all $x,y$. We fix some optimal $I$, call $\KS_I(x \cnd y)$ the conditional complexity of $x$ given $y$, and omit $I$ in the notation (writing just $\KS(x\cnd y)$). The unconditional complexity can be then defined as $\KS(x)=\KS(x\cnd\varepsilon)$, where $\varepsilon$ is the empty string. Both $\KS(x)$ and $\KS(x\cnd y)$ depend on the choice of the optimal $I$, but different choices lead to functions that differ only by an $O(1)$ additive term.

  In our definition $x$ and $y$ are binary strings, but we may use instead arbitrary constructive objects using some computable encoding. The choice of encoding may change the complexity only by an $O(1)$ additive term.
  
  There is another version of complexity where the programs should be ``self-delimited''; formally speaking, we do not allow $I(p,y)$ and $I(p',y)$ to be both defined  if $p$ is a proper prefix of  $p'$. In this restricted class there is also an optimal $I$ that leads to a minimal complexity function; to distinguish it from the function defined above this function is called \emph{prefix complexity}, while the previously defined one is sometimes called \emph{plain complexity}. The difference between plain and prefix complexity is at most $O(\log n)$ for strings of length at most $n$, but for some of our questions this is important. We use plain complexity except for a few places where the use of prefix complexity is mentioned explicitly; letter $\KP$ is used for prefix complexity (and $\KS$ is reserved for plain complexity).
}    

The equation $\KS(x\cnd y)=\KS(x,y)-\KS(y)$ now can be proven (with $O(\log n)$ precision\footnote{We say that two numbers $a$ and $b$ are equal with precision $\Delta$, if $|a-b| \le \Delta$. We also say that $a$ is less than $b$ with precision $\Delta$ if $a \le b + \Delta$.} for $n$-bit strings $x,y$); this proof was found by Levin and Kolmogorov around 1967~\cite{kolmogorov1968} and was one of the first results establishing the connection between Shannon's and algorithmic information theory. The parallelism between Shannon's and algorithmic information theory was already noted in the first publication of Kolmogorov~\cite{kolmogorov1965} whose title was ``Three approaches to the quantitative definition of information'', and the approaches were combinatorial, probabilistic (Shannon's entropy) and algorithmic. This parallelism was developed in many different aspects since then (see~\cite{suv}), but many questions  remain open.

One of the main motivations for Kolmogorov and other inventors of algorithmic information theory was to use the notion of complexity to define individual random objects (and therefore provide foundations for statistics). Martin-L\"of defined randomness in 1966 for infinite sequences; in 1970s Schnorr and Levin established connections between randomness and complexity and now the theory of algorithmic randomness is a well developed theory closely related to algorithmic information theory (see~\cite{nies,downey-hirschfeldt}). Following the tradition of general computability theory, algorithmic randomness deals mostly with infinite objects and qualitative questions (and has obtained very impressive results). This area is mostly out of our scope.  Another interesting topic that is not covered is the time-bounded Kolmogorov complexity and its relation to computational complexity. For this area, see~\cite[Chapter 7]{li-vit:b:kolmbook-four} and~\cite{all:t:mcspsurvey} for an overview of recent developments related to the Minimum Circuit Size Problem.

We survey here several topics in classical Kolmogorov complexity theory centered primarily on finite strings and quantitative aspects, and bring up a number of open questions. We believe that a better understanding of classical Kolmogorov complexity theory may be useful for the two very active subfields mentioned above.  And also for Information Theory in general.

\section{Profiles and inequalities}\label{sec:profiles}

The general theme of this section is: which tuples of numbers can appear as complexities of strings and their combinations? For a given tuple of $k$ strings we consider its \emph{complexity profile} (most of the times we'll call it just \emph{profile}), i.e., $2^{k}-1$ numbers that are complexities of all $2^{k}-1$ non-empty sub-tuples of a given tuple.  For example, for one string its profile is just its complexity; for a pair $(x,y)$ its profile consists of $\KS(x),\KS(y),\KS(x,y)$; a triple of strings $(x,y,z)$ has profile
\[
\big(\KS(x), \KS(y), \KS(z), \KS(x,y), \KS(x,z), \KS(y,z), \KS(x,y,z)\big),
\]
and so on. Obviously, for $k=1$,  every non-negative integer is a complexity of some string (with $O(1)$-precision), so profiles are just positive integers. For $k=2$,  there are some restrictions that correspond to obvious inequalities of monotonicity $\KS(x),\KS(y)\le \KS(x,y)$ and subadditivity  $\KS(x,y)\le \KS(x)+\KS(y)$ that are true with logarithmic precision; these inequalities provide necessary and sufficient conditions for profiles (with the same precision). For $k=3$,  the profile is also described by linear inequalities for complexities; we need to add the submodularity inequality 
\[
\KS(x)+\KS(x,y,z)\le \KS(x,y)+\KS(x,z)
\]
and its variants. 

We do not need to include in the profile the values of \emph{conditional} Kolmogorov complexity since these values can be computed as a combination of the unconditional complexities (e.g., $\KS(x\cnd y) = \KS(x,y) - \KS(y)$, see also Fig.~\ref{fig:venn-xy} and Fig.~\ref{fig:venn-xyz}). 

\def\firstcircleleft{(0,0) circle (1.8cm)}
\def\secondcircleleft{(0:2cm) circle (1.8cm)}
\def\firstcircle{(0,0) circle (1.8cm)}
\def\secondcircle{(0:2cm) circle (1.8cm)}
\def\thirdcircle{(-60:2cm) circle (1.8cm)}

\noindent\begin{minipage}[t]{0.48\linewidth}%
 \hspace*{3em}\raisebox{4em}{
\begin{tikzpicture}[scale=0.9, every node/.style={scale=0.9}]

   \begin{scope}[fill opacity=0.5]
        \fill[black!20] \firstcircle;
        \fill[black!70] \secondcircle;     
         \clip \firstcircle; 
        \fill[black!90] \secondcircle;  
    \end{scope}
   
        \draw[thick] \firstcircle node[below] {};
        \draw[thick] \secondcircle node [below] {};
         \draw  (-2.0,1.5) node  {\scriptsize{$\KS(x)$}};
         \draw  (3.7,1.5) node  {\scriptsize{$\KS(y)$}};
         \filldraw [fill=white, draw=white](-1.30,-0.2) rectangle (-0.30,0.20);
         \draw[ opacity=1.0 ]  (-0.8,0) node  {\scriptsize{$\KS(x\cnd y)$}};
          \filldraw [fill=white, draw=white](2.25,-0.20) rectangle (3.35,0.20);
         \draw[ opacity=1.0 ]  (2.8,0) node  {\scriptsize{$\KS(y\cnd x)$}};
          \filldraw [fill=white, draw=white](1.45,-0.20) rectangle (0.55,0.20);
         \draw[ opacity=1.0 ]  (1.0,0) node  {\scriptsize{$I(x \cln y)$}};
\end{tikzpicture}}


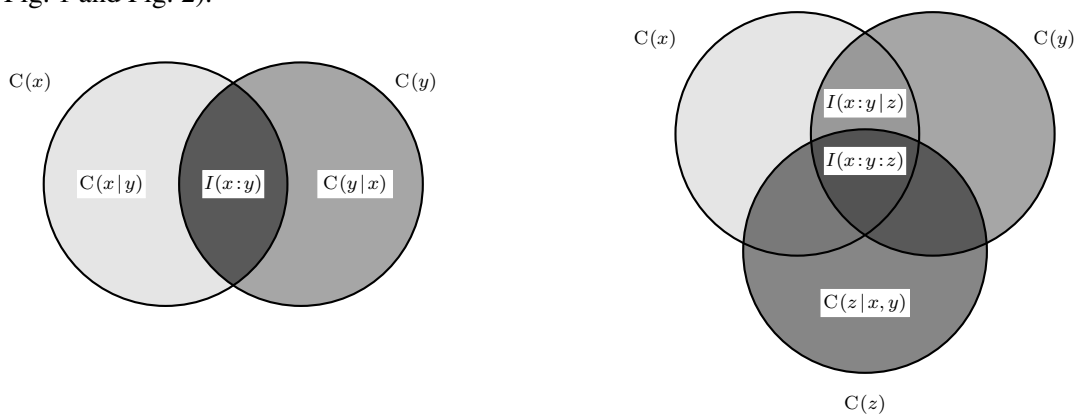
\captionof{figure}{Information quantities for a pair.}\label{fig:venn-xy}
\end{minipage}
 \begin{minipage}[t]{0.48\linewidth}%

\quad\quad\quad\quad
\begin{tikzpicture}[scale=0.9, every node/.style={scale=0.9}]
    \begin{scope}[fill opacity=0.5]
        \fill[black!20] \firstcircle;
        \fill[black!70] \secondcircle;
        \fill[black!95] \thirdcircle;
    \end{scope}

        \draw[thick] \firstcircle node[below] {};
        \draw[thick] \secondcircle node [below] {};
	\draw[thick] \thirdcircle node [below] {};
	
	    \draw  (-2.1,1.4) node  {\scriptsize{$\KS(x)$}};
        \draw  (3.8,1.4) node  {\scriptsize{$\KS(y)$}};
         \draw  (1.0,-4.0) node  {\scriptsize{$\KS(z)$}};
          \filldraw [fill=white, draw=white,  opacity=1.0](0.40,-0.60) rectangle (1.60,-0.20);
         \draw  (1.0,-0.4) node  {\scriptsize{$I(x\cln y\cln z)$}};
         \filldraw [fill=white, draw=white, opacity=1.0](0.35,-2.70) rectangle (1.65,-2.30);
         \draw  (1.0,-2.5) node  {\scriptsize{$\KS(z\cnd x,y)$}};
         \filldraw [fill=white, draw=white, opacity=1.0](0.40,0.25) rectangle (1.60,0.65);
         \draw[ opacity=1.0 ]  (1.0,0.45) node  {\scriptsize{$I(x\cln y\cnd z)$}};
\end{tikzpicture}
\captionof{figure}{Information quantities for a triplet. Warning: the central region can be negative, see the legend.}\label{fig:venn-xyz}
\end{minipage}
 \medskip

\textbf{Legend for \cref{fig:venn-xy}:} 
The area of the left circle is a symbolic representation of $\KS(x)$; the area of the right circle is a symbolic representation of $\KS(y)$. The total area of the union of two circles is the symbolic representation of $\KS(x,y)$, and $\KS(x\cnd y)=\KS(x,y)-\KS(y)$ and $\KS(y\cnd x)=\KS(x,y)-\KS(y)$ with logarithmic precision. Note that we do not claim that the parts can be represented by some strings (see below about the mutual and common information); they just represent some numerical quantities. The middle part has area $I(x\cln y)$ that is (by definition) equal to $\KS(x)+\KS(y)-\KS(x,y)$, or $\KS(x)-\KS(x\cnd y)$, or $\KS(y)-\KS(y\cnd x)$ (all three quantities are the same with logarithmic precision). This quantity measures how much the complexity of $x$ decreases when we get to know $y$, or vice versa; it  is called the amount of \emph{mutual information} between $x$ and $y$.
\smallskip

\textbf{Legend for \cref{fig:venn-xyz}:} Three circles correspond to $x$, $y$, and $z$; the unions of two circles correspond to $\KS(x,y)$, $\KS(x,z)$ and $\KS(y,z)$; the union of all three circles correspond to $\KS(x,y,z)$. This picture has $7$ parts (not counting the infinite surrounding), and we can write $7$ equations for them. One of the equations says that $\KS(x)$ is the sum to four parts inside $x$-circle, two other ones say the same for $y$ and $z$. Then we write an equation saying that $\KS(x,y)$ is the sum of six parts inside $x$- and $y$-circles, and two symmetric equations. The seventh equation says that the sum of all $7$ parts is equal to $\KS(x,y,z)$. It is easy to see that these equations have unique solution, and in this way we introduce new coordinates in the linear space of profiles. Most of the new variables have an intuitive meaning. For example, the bottom part is the conditional complexity of $z$ given $x$ and $y$. The same interpretation works for two other symmetric parts. Three other parts can be interpreted as \emph{conditional mutual information}; for example, $I(x\cln y\cnd z)$ is the mutual information in $x$ and $y$ if $z$ is known, and can be rewritten, e.g., as $\KS(x\cnd z)-\KS(x\cnd y,z)$. The interpretation of the central part is more problematic. We can call it the \emph{mutual information in $x,y,z$}, but should always remember that it could be (unlike the other six parts) negative. For example, this happens if $x$, $y$, $z$ are pairwise independent but not independent: one may take two random independently generated strings for $x$ and $y$ and let $z$ be the bitwise \texttt{xor} of $x$ and $y$.

\medskip

For $k\ge 4$,  the set of all profiles remains unknown. Even the dual question ``which linear inequalities are true for complexity of tuples with logarithmic precision'' (that, roughly speaking, asks only about the convex cone generated by profiles, not about the original set of profiles) is widely open.  It is known that the same linear inequalities are true for entropies of tuples of random variables and for complexities of tuples of strings (with logarithmic precision), and this dual question can be reformulated in terms of Shannon entropies~\cite{hammer2000}, or subgroup sizes~\cite{chan-yeung}, or projection of multi-dimensional sets~\cite{chan-yeung}; see~\cite[Chapter 10]{suv} for details. 
 
For $k\ge 4$, the general structure of the set of  all  linear inequalities for Kolmogorov complexity within logarithmic precision (or, equivalently, for Shannon's entropy)  is not well understood. It is known that these sets of inequalities are not finitely generated (which means that there are infinitely many independent linear information inequalities)~\cite{matus}. However, it remains unknown, for example,  whether these sets are semi-algebraic (which would imply that there are finitely many ``primary'' uniform polynomial inequalities for Kolmogorov complexity and Shannon's entropy, and all other inequalities are corollaries of these ``axioms''). These questions are extensively studied in  information theory, typically in the context of Shannon's entropy and its applications, see the surveys \cite{yeung-survey,chan-survey}.

Thus, the question about profiles goes far beyond the algorithmic information theory. Still there are more specific questions about Kolmogorov complexities related to profiles, and we list some of them in this section. We start with a question that looks very basic but nevertheless is open.

\begin{enumerate}[label=Q\arabic{*}]

\item \label{halve} Let $a_1,\ldots,a_k$ be some strings of length at most $n$.  Is it possible to halve all complexities of $a_i$ and their tuples, i.e., do there exist strings  $a_1',\ldots,a_k'$ such that $\KS(a_i')=0.5\KS(a_i)+O(\log n)$ for all $i$, and $\KS(a_i',a_j')=0.5\KS(a_i,a_j)+O(\log n)$ for all $i$ and $j$, and so on for all sub-tuples? (Here $k$ is assumed to be fixed, and we consider asymptotic results when $n\to\infty$.) 

\end{enumerate}

A similar result for Shannon entropies is known (and rather easy): if $\xi_1,\ldots,\xi_k$ is a tuple of random variables, for every $\varepsilon>0$ there exists another tuple $\xi_1',\ldots,\xi_k'$ such that $H(\xi_i')\approx 0.5H(\xi)$, $H(\xi_i',\xi_j')\approx 0.5 H(\xi_i,\xi_j)$,\ldots,with error at most $\varepsilon$, see~\cite{yeung-textbook}. But the proof of this result looks specific to Shannon entropies.

Of course, there is nothing special in the constant $0.5$, and a similar question can be asked for any other positive factor. We only know that Q1 has positive answer if we replace $0.5$ by $2$ or any other integer factor (the proof uses the same technique of ``typization'' from~\cite{hammer2000} that was used to prove that the same linear inequalities are true for complexities and entropies), and  the question is open for all positive non-integer factors. 

In the question~\ref{halve} the strings $a_i'$ may have no relation with original strings $a_i$. One can additionally require that $\KS(a_i'\cnd a_i)=O(\log n)$ for all~$i$; one can say then that $a_i'$ are obtained from $a_i$ by deleting half of the information. On the other hand, one can let $a_i$ unchanged but ask for an oracle $x$ (a string or an infinite sequence) that halves the complexity of all $a_i$, i.e., $\KS^x (a_i,a_j,\ldots) = 0.5\KS(a_i,a_j,\ldots)$ for all $i,j,\ldots$. Both claims imply the original one. For the first claim this is obvious; for the second one we may note that the set of possible profiles does not change if we relativize complexity function to an arbitrary oracle; this can be proven using the ``typization technique'' from~\cite{hammer2000} (see also~\cite[Chapter 10]{suv} for details).
\medskip

\noindent
We now move to questions related to the fact, already mentioned above, that exactly the same classes of linear inequalities are valid for Kolmogorov complexity (with a logarithmic precision) and for Shannon's entropy.
However, for entropies we may also consider \emph{constraint} inequalities that say that if some equation for entropies is true (e.g., if $I(X\cln Y\cnd Z)=0$), then some inequality $A(X,Y,Z,\ldots)\le 0$ is true (where $A$ is some linear combination of entropies of $X,Y,Z,\ldots$ and their tuples). Some of these inequalities have ``non constraint versions'': for example, if we manage to prove (unconditionally) that $A(X,Y,Z,\ldots)\le cI(X\cln Y\cnd Z)$ for some constant $c$, then the conditional inequality is guaranteed to be true. However, it is known that for some conditional inequalities this is not the case~\cite{kaced-romashchenko}, and it is not clear what could be the combinatorial translation of those conditional inequalities. (It seems that there might be even subtler differences between different constraint inequalities, see a classification of the known constraint inequalities for Shannon entropy and for Kolmogorov complexity in~\cite{kaced-romashchenko}.)

The equivalence result says that if some linear inequality is true for Shannon entropies, it is also true for Kolmogorov complexities with logarithmic precision (up to $O(\log n)$ additive term for strings of length at most $n$). However, in many cases the inequality for complexities becomes true with maximal possible precision $O(1)$ if we switch to \emph{prefix} complexities. The prefix complexity $\KP(x)$ is a special version of complexity that is defined using prefix-free machines and self-delimited programs, see, e.g., \cite{downey-hirschfeldt,suv}. We do not go into details of its definition, noting only that $\KP(x)$ coincides with $\KS(x)$ with $O(\log n)$ precision for strings of length at most $n$. If we replace $\KS$ by $\KP$, the submodularity inequalities become valid with $O(1)$-precision. Therefore, the same happens for all their linear combinations (the so-called Shannon inequalities).

\openq{Do all linear inequalities for entropies (or complexities, since they are the same), or at least the known non-Shannon inequalities, hold with precision $O(1)$ for the case of prefix complexity?}

\medskip

\openq{
The combinatorial translations of inequalities for entropies and complexities~\cite{combinatorial2000,suv} also include some additional factors (polynomial in the cardinalities) that correspond to logarithmic additive terms for the complexity inequalities. Again, for many examples (including the subadditivity and submodularity) these polynomial factors are unnecessary. Can we get rid of these polynomial factors in the general case?
}
\medskip

\noindent 
The next question is about relativization. There are two types of relativization that are possible for Kolmogorov complexity. We can fix some infinite sequence of zeros and ones (or a set of natural numbers) and add it as an oracle in all definitions of algorithmic information theory. Or we can take some finite string and add it as a condition everywhere. The first approach seems more general (any finite string may be converted to an equivalent oracle), but how much more general is it? This question can be formalized as follow. Fix some infinite sequence $X$. Then, for every $n$, we may consider the complexities of all strings of length at most $n$ with oracle $X$.

\openq{Does there exist a finite string $x_0$ such that the conditional complexities of all strings (of length at most $n$) with condition $x_0$ will coincide with $O(\log n)$ precision with $X$-relativized complexities? If yes, how long does this string $x_0$ need to be? Can we, for example, always find such a string $X_0$ of length $\poly(n)$?
}

One may ask the same question not for all strings of length at most $n$, but for polynomially many of them: Is it true that for any oracle $X$, for any $n$ and for any set $S$ that contains polynomially (in $n$) many strings of size at most $n$, there is a string $x_0$ such that the oracle $X$ changes the complexities of the elements of $S$ in the same way as the condition $x_0$, up to $O(\log n)$ precision? 


\section{Information-theoretic properties}
\label{sec:info-theoretic}

This rather vague title means, informally speaking, that we are interested in the properties of finite objects that can be expressed in the language of the information theory. For example, one can observe that information contained in every string $x$ is ``divisible'' in the following sense: one can find two strings $x_1$ and $x_2$ of complexity $\KS(x)/2$ that together contain all the information in $x$, i.e., $\KS(x\cnd x_1,x_2)=0$ (all with logarithmic precision). This is an information-theoretic result: we do not care whether strings $x,x_1,x_2$ start with $0$ or $1$, what exactly their length is, etc., (but care about their complexity).

Then some general and not quite well-defined question arises: is the complexity of a string $x$ its \emph{only} information-theoretic characteristic? Can it happen that two strings $x$ and $x'$ with the same complexity have drastically different information-theoretic behavior? Here is one way to formalize this question:

\openq{
For a given string $x$ and given $k$, we consider the set of profiles of $(k+1)$-tuples $(x, y_1, \ldots, y_k)$ for all $k$-tuples $y_1,\ldots,y_k$.  This set depends only on $x$; is it true that this set (for given $k$), considered with logarithmic precision, depends only on the \emph{complexity} of $x$?
}

We may equivalently ask this question for all random (incompressible) strings of a given length. Indeed, every string $x$ of complexity $k$ is ``informationally close'' to some random string of length $k$, namely, to the shortest program for $x$. Here two strings $x$ and $y$ are called \emph{informationally close}, if the information distance
\[  
  d(x,y) = \KS(x\cnd y)+\KS(y\cnd x)
\]  
between them is small. It is easy to see that the set mentioned in our question is almost the same for informationally close string. So, if our question has a positive answer (complexity is the only important information-theoretic property), this fact can be considered as a finite analogue of Kolmogorov's $0$-$1$-law for invariant properties of random sequences.

Another way of distinguishing strings of the same complexity can be the following one. Given a string $x$ of complexity $n$, and some $m<n$, we can study the axiomatic power of the statement $\KS(x)>m$, in particular, the complexity of the universal statements that follow from it. (Universal statements are statements about non-termination of some program in an optimal programming language, and their complexity is the length of this program, see~\cite{bienvenu-axioms} for details.) If $x$ and $x'$ have small information distance, then we can positively check this, and therefore the high complexity of $x$ provably implies (almost as) high complexity of $x'$ and vice versa. So the axiomatic power of a string (in the sense explained above) is an invariant property.

\openq{Does the axiomatic power distinguish strings of the same complexity?}

\emph{Remark.} In fact, algorithmic statistics does distinguish strings of the same complexity using more subtle properties of strings like the \emph{Kolmogorov structure function} that (for a given string $x$) is defined as
\[
k\mapsto \text{minimal log-cardinality of a finite set of complexity at most $k$ that contains $x$};
\]
note that here we speak not only about complexity of strings, but also about complexity of finite sets of strings (defined as the complexity of their encodings).

Kolmogorov structure function is well studied in algorithmic statistics; it can be defined in several equivalent ways (resource-bounded complexity, $(\alpha,\beta)$-stochacticity, computational/logical depth, see, e.g., the survey~\cite{algorithmic-statistics-survey} for details and historical information).

However, this is a rather subtle property of a string $x$ that can be drastically different for two strings $x$ and $x'$ that are informationally close  (in terms of algorithmic statistics, one of them can be stochastic while there other is far from being stochastic). If we are interested in this kind of properties, we should modify the definition of informationally close strings using the total complexity $\KT(x\cnd y)$ defined as the length of a minimal \emph{total} program that maps $y$ to $x$. But this goes beyond the scope of our survey.

\section{Extracting common information}\label{s:extractci}

When we switch from individual strings to pairs of strings, we get a much more diverse landscape: the complexity profile of a pair $(x,y)$, i.e., a triple $(\KS(x),\KS(y),\KS(x,y))$, does not determine the information-theoretic properties of the pair $(x,y)$. This was found long ago by G\'acs and K\"orner who realized that there are pairs with the same profile  and for one pair the ``mutual information'' $I(x\cln y)=\KS(x)+\KS(y)-\KS(x,y)$ is extractable and for the other one  it is not~\cite{gacs-korner}. Extractability means that the diagram of Figure~\ref{fig:venn-xy} can be understood literally:
there are three strings $(p,q,r)$ that correspond to the three parts of the diagram. This means that
\begin{itemize}
    \item $p,q,r$ are independent (have mutual information close to zero);
    \item $x$ is informationally close to $(p,q)$;
    \item $y$ is informationally close to $(q,r)$.
\end{itemize}
Then we can say that $x$ and $y$ have common information $q$, while $x$ (in addition to the common part $q$) has some private part $p$, and $y$ (in addition to $q$) has some private part $r$. It is easy to see that in this case complexities of $p$, $q$ and $r$ respectively are
\(
  \KS(x\cnd y), \ I(x\cln y), \ \KS(y\cnd x),
\)
and that $\KS(q\cnd x)$ and $\KS(q\cnd y)$ are close to~$0$.

A second way to define the extractability of mutual information: there exists a string $q$ of complexity close to $I(x\cln y)$ that is simple relative to both $x$ and $y$. Then $p$ and $r$ can be chosen as the minimal programs to produce $x$ and $y$ given $q$. 
 
And a third way to define extractability: there exists a string $q$ of complexity $I(x\cln y)$ that makes $x$ and $y$ independent, i.e.,  $I(x\cln y\cnd q)=0$. 
 
It is easy to check that all these definitions are equivalent, and, furthermore, if a string $q$ certifies extractability of mutual information in one way, then it also certifies it in the other two. In a very intuitive sense (or rather, three senses) the string $q$ represents the \emph{common information} in $x$ and $y$. 

To summarize: the mutual information $I(x\cln y)$ is just a number, but for some $x$ and $y$ this mutual information can be ``extracted,'' or ``materialized'' in the form of a string, and then this string is called the common information in $x$ and $y$.

It is easy to find strings where common information can be extracted. Just take independent $p, q, r$ and let $x=(p,q)$ and $y=(q,r)$. Proving non-extractability is a more difficult task. G\'acs and K\"orner proved~\cite{gacs-korner} that the repeated sampling of dependent random variables leads (with high probability) to two strings that do not have extractable mutual information, except for some trivial cases. Many other examples of pairs with non-extractable mutual information were constructed later. In particular, Muchnik showed that for a random pair of incident line and point on a finite affine plane those line and point have non-extractable mutual information~\cite{muchnik-upper-semilattice}. Still quantitatively this phenomenon is not well understood.

Quantitative understanding means that we want to measure how far is the pair $(x,y)$ from the situation we described above (that can be called ``full extractability''). Different equivalent definitions of extractability lead to different measures.

\subsubsection*{Gray--Wyner profile}

For strings $(x,y)$ with (fully extractable) common information there exists a string $q$ such that $\KS(q)=I(x\cln y)$, $\KS(x\cnd q)=\KS(x\cnd y)$, and $\KS(y\cnd q)= \KS(y\cnd x)$. This motivates the following definition: for a given pair of strings $(x,y)$ we consider all triples of integers $(\alpha,\beta,\gamma)$ for which there exists some $q$ with $\KS(q)\le\gamma$, $\KS(x\cnd q)\le\alpha$ and $\KS(y\cnd q)\le \beta$. We call this set the \emph{Gray--Wyner profile} of the pair $(x,y)$; the Shannon's version of this notion for independent samples of random variables was considered in~\cite{gra-wyn:j:gwcode}. Fully extractability of common information means that Gray--Wyner profile contains the triple $(I(x\cln y), \KS(x\cnd y), \KS(y\cnd x))$ (up to logarithmic terms, as usual). In fact, for strings with extractable common information the Gray--Wyner profile is maximal. We also know the tight \emph{lower} bound for the Gray--Wyner profile, i.e., we can give an example of a pair where the Gray--Wyner profile is minimal~\cite[Chapter 11]{suv}.

\openq{\label{randlinepoint}  Determine the Gray--Wyner profiles for specific examples of pairs where the mutual information is not extractable (e.g., for a random pair of incident line and point on the affine plane and for G\'acs--K\"orner example; in the latter case one may hope to adapt the results from~\cite{gra-wyn:j:gwcode}). Describe all possible Gray--Wyner profiles for pairs of strings. (The same questions can be asked for Wyner profile and extended profile defined below.)}

\subsubsection*{Wyner profile}
As we mentioned above, for a pair $(x,y)$ with (fully extractable) common information one can find a string $q$ of complexity $I(x\cln y)$ that makes $x$ and $y$ independent, i.e., $I(x\cln y\cnd q)$ is (almost) zero. If we do not bound the complexity of $q$, such a string (that makes $x$ and $y$ independent) always exists, we may let $q=x$ or $q=y$. The question is how much we can decrease the complexity of $q$ while keeping $I(x\cln y\cnd q)$ reasonably small. This can be formalized as follows: the \emph{Wyner profile} of the pair $(x,y)$ is the set of all $\alpha$ and $\beta$ such that there exists some $q$ such that $\KS(q)\le \alpha$ and $I(x\cln y\cnd q)\le \beta$. An analogous question for Shannon's entropy was raised by A.~Wyner in the classical paper \cite{wyner1975}.

\subsubsection*{Extended profile}
There are other ways to measure the extractability of common information, e.g., one may look at the maximal complexity of $q$ such that both $\KS(q\cnd x)$ and $\KS(q\cnd y)$ do not exceed some $\alpha$. All these measures can be derived from the most informative one that can be called the \emph{extended profile} and is defined as follows: consider all strings $q$ and for each of them consider the profile of the \emph{triple} $(x,y,q)$. The set of all these profiles is called the \emph{extended profile} of the pair $(x,y)$. 

A profile of a triple is a point in $7$-dimensional set, but for fixed $x$ and $y$ we have three quantities (that form the profile of a pair) fixed, so essentially the extended profile is a four-dimensional set. Note that the Gray--Wyner profile was three-dimensional, and the Wyner profile was two-dimensional. So a natural question arises:

\openq{\label{wynerprofile} Does the extended profile contain more information about $(x,y)$ than the Gray--Wyner profile, or Wyner profile? Is it possible to find two pairs that have (almost) the same Gray--Wyner or Wyner profiles but significantly different extended profiles? Does the Gray--Wyner profile determine the Wyner profile? Does the Wyner profile determine the Gray--Wyner profile? Do we get even more information if we consider a tuple of strings $(q_1,\ldots,q_k)$ instead of one string $q$ in the definition of the extended profile?
} 

As we have mentioned, all the notions (and most results) of the algorithmic information theory can be relativized by adding some oracle to all computations.
It is easy to see that relativizing the complexity function by \emph{any} oracle does not increase the complexity of a given string. On the other hand, relativizing the complexity function by a \emph{random} oracle does not decrease the complexity of a given string with probability close to $1$.  So, a random oracle does not change the complexity profile of given tuple of strings. However, it is not obvious whether the same is true for more complicated information-theoretic properties of strings. Here is one natural question of this type.

\openq{\label{ro}Is it true that \emph{random} oracles do not change the Gray--Wyner profile with high probability? A similar question can be asked for other types of profiles we considered.
}

In  \cite{muchnik-stability} a positive answer to this question was proven for stochastic tuples of strings, using the restricted algorithmic Ahlswede--K\"orner lemma, which takes us to the next section. 

\section{The Ahlswede--K\"orner lemma and Kolmogorov complexity}\label{s:profile-reduction}

It was suggested in \cite{muchnik-stability} that~\ref{ro}  is linked with the   (seemingly technical) problem that we bring up in this section. Actually, the questions~\ref{ak} and~\ref{akexp} below are stand-alone problems that are 
interesting in their own right. These questions are also connected with a more general problem of extraction of common information (see the discussion on this type of problems in~\cref{s:srp2}). 

\openq{\label{ak}Is it true that for every pair  of strings $(x,y)$ of complexity at most $n$ and for every string $z$ there exists another string $z'$ such that the complexity values 
\begin{equation}\label{e:ak1}
\big(\KS(x\cnd z), \KS(y\cnd z), \KS(x,y\cnd z)\big)
\end{equation}
are equal (with a logarithmic precision) to 
\begin{equation}\label{e:ak2}
\big(\KS(x\cnd z'), \KS(y\cnd z'), \KS(x,y\cnd z')\big)
\end{equation}
and at the same time $\KS(z'\cnd x,y) = O(\log n)$? 
}

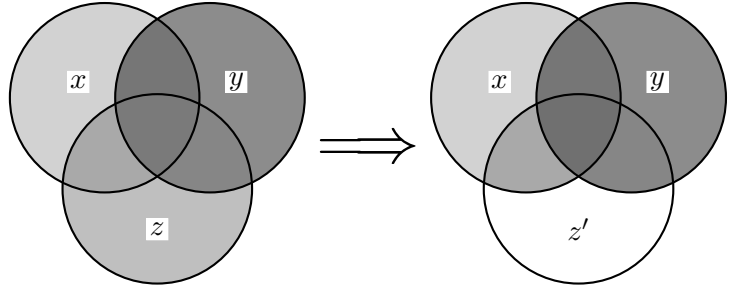
\begin{figure}[h]
\begin{center}
\begin{tikzpicture}[scale=0.7]
    \begin{scope}[fill opacity=0.5]
        \fill[black!35] \firstcircle;
        \fill[black!90] \secondcircle;
        \fill[gray] \thirdcircle;

        \draw[thick] \firstcircle node {};
        \draw[thick] \secondcircle node  {};
	    \draw[thick] \thirdcircle node  {};

        \filldraw [fill=white, draw=white, opacity=1.0](-0.7,0.1)  rectangle (-0.3,0.5);
        \draw[thick, opacity=1.0] (-0.5,0.3)  node {$x$};
        \filldraw [fill=white, draw=white, opacity=1.0](2.3,0.1)  rectangle (2.7,0.5);
        \draw[thick, opacity=1.0] (2.5,0.3) node  {$y$};
        \filldraw [fill=white, draw=white, opacity=1.0](0.8,-2.7)  rectangle (1.2,-2.3);
	    \draw[thick, opacity=1.0] (1.0,-2.5) node {$z$};
    \end{scope}

 \begin{scope}[shift={(8,0)}]
 
	\fill[white] \thirdcircle;

    \begin{scope}[fill opacity=0.5]
    	\clip \firstcircle;
        \fill[black!35] \firstcircle;
        \fill[gray] \thirdcircle;
    \end{scope}

   \begin{scope}[fill opacity=0.5]
    	\clip \secondcircle;
        \fill[black!90] \secondcircle;
        \fill[gray] \thirdcircle;
    \end{scope}
        
        \draw[thick] \firstcircle node {};
        \draw[thick] \secondcircle node {};
	\draw[thick] \thirdcircle node  {};

        \filldraw [fill=white, draw=white, opacity=1.0](-0.7,0.1)  rectangle (-0.3,0.5);
        \draw[thick, opacity=1.0] (-0.5,0.3)  node {$x$};
        \filldraw [fill=white, draw=white, opacity=1.0](2.3,0.1)  rectangle (2.7,0.5);
        \draw[thick, opacity=1.0] (2.5,0.3) node  {$y$};
	    \draw[thick] (1.0,-2.5) node {$z'$};

\end{scope}

\draw[thick]  (5,-1) node {\Huge $\Longrightarrow$};
\end{tikzpicture}
\end{center}
\caption{Question~\ref{ak} in a picture. Six areas of the same color (on the left and on the right) represent equal complexity quantities. The vanishing information quantity is shown in white.}\label{fig:venn-ahslwede-koerner}
\end{figure}

We do not require in this question that $z'$ is simple relative to $z$ as well, it would be too much: in the case $x=y$ this would require extracting the mutual information from $x$ and $z$.

This statement is the algorithmic information theory version of the Ahlswede--K\"orner lemma in the Shannon information theory (where it is proven for independent samples of random variables, see \cite{AhlswedeKorner}). There is also a partial result for the algorithmic information theory version: the claim is true if the pair $(x,y)$ is stochastic (i.e., is an element of almost maximal complexity in a simple finite set, see \cite[Section~14.2]{suv}). This result can be partially constructivized: the string $z'$ can be computed given $(x,y)$, the profile of $(x,y)$, by a double exponential time algorithm (that also gets oracle access to the set that appears in the definition of the stochasticity).

\openq{\label{akexp} Can the double exponential runtime be improved?}

\section{Communication problems: Shannon and algorithmic version}\label{s:comm}

 Multi-source network coding problems  have been extensively studied in the context of  Shannon's information theory. In these problems we are given a network (a directed  graph), where some nodes are associated with input \emph{sources} (mutually correlated random variables, usually with some properties of ergodicity),  and some nodes are \emph{terminals}, which wish to reconstruct specific parts of the input data. The edges play the role of communication channels with limited capacities. The aim is to establish the relation between the entropy profile of the input  sources and the capacities of the edges that allow vertices to perform encoding/decoding satisfying the capacity constraints and allow the terminal nodes to obtain the desired data.
 
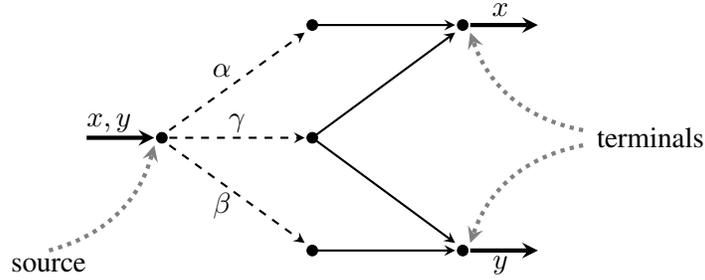
\begin{figure}
\begin{center}
\begin{tikzpicture} 

\filldraw  (0,0) circle (2pt) ;
\filldraw  (2,0) circle (2pt) ;

\filldraw  (2,1.5) circle (2pt) ;
\filldraw  (2,-1.5) circle (2pt) ;

\filldraw  (4,1.5) circle (2pt) ;
\filldraw  (4,-1.5) circle (2pt) ;

\draw[ ultra thick,-stealth] (-1.0,0) -- (-0.1,0);

\draw[ thick,-stealth] (2,1.5) -- (3.9,1.5);
\draw[ thick,-stealth] (2,-1.5) -- (3.9,-1.5);

\draw[ thick,-stealth] (2,0) -- (3.9,1.4);
\draw[ thick,-stealth] (2,0) -- (3.9,-1.4);

\draw[ ultra thick,-stealth] (4.1,1.5) -- (5.0,1.5);
\draw[ ultra thick,-stealth] (4.1,-1.5) -- (5.0,-1.5);

\draw[ dashed, thick,-stealth] (0.1,0.1) -- (1.9,1.4);
\draw[ dashed, thick,-stealth] (0.1,-0.1) -- (1.9,-1.4);
\draw[ dashed, thick,-stealth] (0.1,0) -- (1.9,0);

\draw  (-0.7,0.2) node {$x,y$} ;
\draw  (4.5,1.7) node {$x$} ;
\draw  (4.5,-1.7) node {$y$} ;

\draw  (0.8,0.9) node {$\alpha$} ;
\draw  (0.8,-0.9) node {$\beta$} ;
\draw  (1.0,0.2) node {$\gamma$} ;

\draw  (-1.5,-1.7) node {source} ;
\draw[ dotted, gray, ultra thick,-stealth,bend right] (-1.5,-1.5) to (-0.1,-0.1);

\draw  (6.5,0.0) node {terminals} ;
\draw[ dotted, gray, ultra thick,-stealth,bend left]  (5.6,0.1) to (4.1,1.4);
\draw[ dotted, gray, ultra thick,-stealth,bend right]  (5.6,-0.1) to (4.1,-1.4);

\end{tikzpicture}

\end{center}
 \caption{Network information transmission problem that corresponds to Gray--Wyner profile.}
\label{f:graywyner}
\end{figure}

Such problems can be studied in the framework of algorithmic information theory as well,  where the data are represented by strings, and their correlations are measured in terms of Kolmogorov complexity. One specific example, motivated by the problem of \emph{Gray--Wyner profile} discussed above, is shown in~\cref{f:graywyner}. At the \emph{source} (the leftmost vertex) both strings $x$ and $y$ are known. Three channels of bounded capacity (the dotted lines) connect the source with three intermediate nodes, and then the unlimited capacity channels (the solid lines) connect intermediate nodes with terminals (two vertices on the right) that should output messages $x$ and $y$ (respectively).
For this network we ask whether there exist some messages on all edges that have complexity bounded by the edge capacity such that no new information is created at every vertex (the outgoing messages are simple given the incoming ones). There are three edges of bounded capacity, so we look for three messages: $q$ of complexity at most $\gamma$ for the middle edge, and $p$ and $r$ of capacities at most $\alpha$ and $\beta$ for the top and bottom edge. (The edges of unbounded capacity may just transmit all the available information.) The messages $p$ and $q$ together should be enough to reconstruct $x$, so $\KS(x\cnd q)\le \alpha$; for the same reasons, $\KS(y\cnd q)\le \beta$. So, if the information transmission problem is solvable, the triple $(\alpha,\beta,\gamma)$ belongs to the Gray--Wyner profile. The reverse statement is also true: if there is some $q$ such that $\KS(q)\le \gamma$, $\KS(x\cnd q)\le \alpha$, and $\KS(y\cnd q)\le \beta$, then we can let $p$ and $q$ be the shortest programs that transform $q$ to $x$ and $y$, respectively. The only missing conditions are that $p,q,r$ are simple given $x$ and $y$ (the information transmission problem requires that no new information appears in the leftmost vertex). If this is not the case, we can replace $p,q,r$ by other strings that have the required properties (have correct lengths and are enough to reconstruct  $x$ and $y$) found by search; note that the properties are computably enumerable, so we can wait until the first triple appears, and this triple will be simple given $x$ and $y$. We reformulated the notion of Gray--Wyner profile in terms of a multi-source information problem. For each network there are simple conditions corresponding to the information flow through each cut of the network that are necessary for satisfying the constraints (derived as in the max-flow min-cut theorem, see~\cite[Chapter 12]{suv} for the details). 

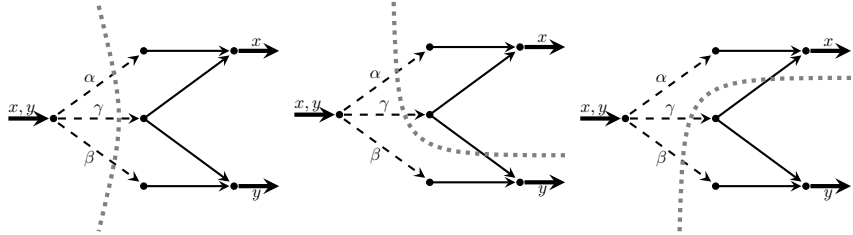
\begin{figure}
\begin{center}
\begin{tikzpicture}[scale=0.6, every node/.style={scale=0.6}] 
\filldraw  (0,0) circle (2pt) ;
\filldraw  (2,0) circle (2pt) ;

\filldraw  (2,1.5) circle (2pt) ;
\filldraw  (2,-1.5) circle (2pt) ;

\filldraw  (4,1.5) circle (2pt) ;
\filldraw  (4,-1.5) circle (2pt) ;

\draw[ ultra thick,-stealth] (-1.0,0) -- (-0.1,0);

\draw[ thick,-stealth] (2,1.5) -- (3.9,1.5);
\draw[ thick,-stealth] (2,-1.5) -- (3.9,-1.5);

\draw[ thick,-stealth] (2,0) -- (3.9,1.4);
\draw[ thick,-stealth] (2,0) -- (3.9,-1.4);

\draw[ ultra thick,-stealth] (4.1,1.5) -- (5.0,1.5);
\draw[ ultra thick,-stealth] (4.1,-1.5) -- (5.0,-1.5);

\draw[ dashed, thick,-stealth] (0.1,0.1) -- (1.9,1.4);
\draw[ dashed, thick,-stealth] (0.1,-0.1) -- (1.9,-1.4);
\draw[ dashed, thick,-stealth] (0.1,0) -- (1.9,0);

\draw  (-0.7,0.2) node {$x,y$} ;
\draw  (4.5,1.7) node {$x$} ;
\draw  (4.5,-1.7) node {$y$} ;

\draw  (0.8,0.9) node {$\alpha$} ;
\draw  (0.8,-0.9) node {$\beta$} ;
\draw  (1.0,0.2) node {$\gamma$} ;

\draw[gray,dotted,ultra thick] (1.0, 2.5) .. controls (1.6, 0.0) .. (1.0, -2.5);
\end{tikzpicture}
\begin{tikzpicture}[scale=0.6, every node/.style={scale=0.6}] 
\filldraw  (0,0) circle (2pt) ;
\filldraw  (2,0) circle (2pt) ;

\filldraw  (2,1.5) circle (2pt) ;
\filldraw  (2,-1.5) circle (2pt) ;

\filldraw  (4,1.5) circle (2pt) ;
\filldraw  (4,-1.5) circle (2pt) ;

\draw[ ultra thick,-stealth] (-1.0,0) -- (-0.1,0);

\draw[ thick,-stealth] (2,1.5) -- (3.9,1.5);
\draw[ thick,-stealth] (2,-1.5) -- (3.9,-1.5);

\draw[ thick,-stealth] (2,0) -- (3.9,1.4);
\draw[ thick,-stealth] (2,0) -- (3.9,-1.4);

\draw[ ultra thick,-stealth] (4.1,1.5) -- (5.0,1.5);
\draw[ ultra thick,-stealth] (4.1,-1.5) -- (5.0,-1.5);

\draw[ dashed, thick,-stealth] (0.1,0.1) -- (1.9,1.4);
\draw[ dashed, thick,-stealth] (0.1,-0.1) -- (1.9,-1.4);
\draw[ dashed, thick,-stealth] (0.1,0) -- (1.9,0);

\draw  (-0.7,0.2) node {$x,y$} ;
\draw  (4.5,1.7) node {$x$} ;
\draw  (4.5,-1.7) node {$y$} ;

\draw  (0.8,0.9) node {$\alpha$} ;
\draw  (0.8,-0.9) node {$\beta$} ;
\draw  (1.0,0.2) node {$\gamma$} ;

\draw[gray,dotted,ultra thick] (1.2, 2.5) .. controls (1.3, -0.9)   .. (5.0, -0.9);

\draw (-1, -2.5)  node {} ; 

\end{tikzpicture}
\begin{tikzpicture}[scale=0.6, every node/.style={scale=0.6}] 
\filldraw  (0,0) circle (2pt) ;
\filldraw  (2,0) circle (2pt) ;

\filldraw  (2,1.5) circle (2pt) ;
\filldraw  (2,-1.5) circle (2pt) ;

\filldraw  (4,1.5) circle (2pt) ;
\filldraw  (4,-1.5) circle (2pt) ;

\draw[ ultra thick,-stealth] (-1.0,0) -- (-0.1,0);

\draw[ thick,-stealth] (2,1.5) -- (3.9,1.5);
\draw[ thick,-stealth] (2,-1.5) -- (3.9,-1.5);

\draw[ thick,-stealth] (2,0) -- (3.9,1.4);
\draw[ thick,-stealth] (2,0) -- (3.9,-1.4);

\draw[ ultra thick,-stealth] (4.1,1.5) -- (5.0,1.5);
\draw[ ultra thick,-stealth] (4.1,-1.5) -- (5.0,-1.5);

\draw[ dashed, thick,-stealth] (0.1,0.1) -- (1.9,1.4);
\draw[ dashed, thick,-stealth] (0.1,-0.1) -- (1.9,-1.4);
\draw[ dashed, thick,-stealth] (0.1,0) -- (1.9,0);

\draw  (-0.7,0.2) node {$x,y$} ;
\draw  (4.5,1.7) node {$x$} ;
\draw  (4.5,-1.7) node {$y$} ;

\draw  (0.8,0.9) node {$\alpha$} ;
\draw  (0.8,-0.9) node {$\beta$} ;
\draw  (1.0,0.2) node {$\gamma$} ;

\draw[gray,dotted,ultra thick] (1.2, -2.5) .. controls (1.3, 0.9)   .. (5.0, 0.9);

\draw (-1, 2.5)  node {} ; 

\end{tikzpicture}
\end{center}
 \caption{Three cuts provide three necessary conditions for the points in the Gray--Wyner profile.}
\label{f:graywyner2}
\end{figure}

For example, for the Gray--Wyner network one can consider three cuts (Figure~\ref{f:graywyner2}).
The first one says that $p,q,r$ together should provide enough information to reconstruct $x,y$, so the triple $(\alpha,\beta,\gamma)$ can be in the Gray--Wyner profile only if $\alpha+\beta+\gamma\ge \KS(x,y)$. The second and third cuts show that $\alpha+\gamma\ge \KS(x)$ and $\beta+\gamma\ge\KS(y)$ for every point in the Gray--Wyner profile.

We get some upper bound for the Gray--Wyner profile, and this upper bound is reached if the common information exists. For the cases where the common information does not exist, the Gray--Wyner profile is smaller, and these necessary conditions are no longer sufficient.

\begin{figure}
\begin{center}
\begin{tikzpicture} 

\draw[ ultra thick,-stealth] (1.0,0) -- (1.9,0);
\draw[ thick, dashed, -stealth ] (2.1,0) -- (3.9,0);
\draw[ ultra thick,-stealth] (4,0) -- (5.0,0);

\draw[ ultra thick,-stealth] (3,-1) -- (3.9,-0.1);

\filldraw  (2,0) circle (2pt) ;
\filldraw  (4,0) circle (2pt) ;

\draw  (1.5,0.2) node {$x$} ;
\draw  (4.5,0.2) node {$x$} ;
\draw  (3.6,-0.7) node {$y$} ;

\end{tikzpicture}

\end{center}
 \caption{Network information transmission problem for the Muchnik conditional codes theorem.}
\label{f:muchnik}
\end{figure}
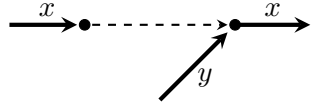

For some other networks the information flow conditions are not only always necessary but also always sufficient. One of the first examples was Muchnik's theorem on conditional codes~\cite{muchnik-conditional} that deals with the network in~\cref{f:muchnik}.
Here the terminal needs to reconstruct $x$ knowing $y$, and gets some information about $x$ through the bounded channel; how much information does it need? In other words, we are looking for a program $p$ that is simple conditional on a string $x$ and that produces $x$ given  a string $y$ as a condition.  Obviously, the size of such a program cannot be less than $\KS(x \cnd y)$, which is by definition the length of the shortest program that transforms $y$ (known to the receiver) to $x$. The difficulty is that the sender does not know $y$ (if it knew $y$, then such a program could be found by search: we need only to know its length, and this is logarithmic amount of information).

\label{p:docexchange} This problem has the following interpretation. Suppose that Alice wants to send a document $x$ to Bob, who holds an older version $y$ not known by Alice. Then Alice would like to get a short description of $x$ given $y$ (something like a \emph{diff} file), without having $y$.\footnote{This problem is known as \emph{document exchange} and also as \emph{information reconciliation} in the theoretical computer science (TCS) literature, and as \emph{asymmetric Slepian--Wolf coding} and also as \emph{compression with side information at the receiver} in the information theory (IT) literature. Typically the scenario in TCS is that the Hamming distance or the edit distance between $x$ and $y$ is small, and in IT that $x$ and $y$ are realizations of random variables $X$ and $Y$ with small entropy $H(X \cnd Y)$. The corresponding scenario in Kolmogorov complexity (namely $\KS(x \cnd y)$ is small) is arguably the least restrictive one in this group.} Muchnik's theorem demonstrates that, surprisingly, the absence of $y$ does not cost too much the sender:  the ``overhead'' is only logarithmic. That is, there exists  a program $p$ of length  $\KS(x \cnd y)+O(\log \KS(x,y))$ producing $x$ conditioned by $y$   such that $\KS(p \cnd x) = O(\log \KS(x,y))$. (In the next section we shall see that  in some circumstances the  program $p$ can be found efficiently.)  Thus, the capacity of the bounded channel in~\cref{f:muchnik} is necessarily at least $\KS(x\cnd y)$, and Muchnik's theorem guarantees that  this capacity is essentially sufficient.

\begin{figure}
\begin{center}
\begin{tikzpicture} 

\draw[ ultra thick,-stealth] (1.0,1) -- (1.9,1);
\draw[ ultra thick,-stealth] (1.0,-1) -- (1.9,-1);

\draw[ thick, dashed, -stealth ] (2.1,1) -- (3.9,0.1);
\draw[ thick, dashed, -stealth ] (2.1,-1) -- (3.9,-0.1);

\draw[ ultra thick,-stealth] (4,0) -- (5.0,0);

\filldraw  (2,1) circle (2pt) ;
\filldraw  (2,-1) circle (2pt) ;

\filldraw  (4,0) circle (2pt) ;

\draw  (1.5,1.2) node {$x$} ;
\draw  (1.5,-0.8) node {$y$} ;

\draw  (4.5,0.2) node {$x,y$} ;

\draw  (3.0,0.8) node {$\alpha$} ;
\draw  (3.0,-0.8) node {$\beta$} ;

\end{tikzpicture}
\end{center}
 \caption{Network information transmission problem for the Slepian--Wolf theorem.}
\label{f:slepian-wolf}
\end{figure}
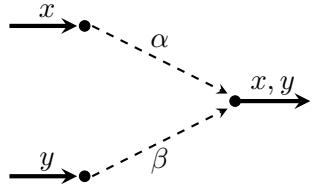

Another network for which the information flow conditions are always sufficient, is the Slepian--Wolf network in~\cref{f:slepian-wolf}; it was considered in~\cite{Slepian--Wolf} in the Shannon information theory setting; the algorithmic information theory version was studied in~\cite{rom:j:slepwolf} (English version~\cite {rom:t:slepwolf}),\cite{zim:c:kolmslepianwolf}, \cite{bau-zim:t:univcompression}.
Here the receiver wants to know both strings $x$ and $y$, and two senders know $x$ and $y$ respectively. The first is allowed to send $\alpha$ bits, while the second is allowed to send $\beta$ bits. For which $\alpha$ and $\beta$ is this possible? Note that for unbounded $\beta$ we get the previous information transmission problem.

There are information flow bounds: 
\begin{itemize}
\item $\alpha+\beta\ge \KS(x,y)$, since the receiver should reconstruct the pair $(x,y)$ having only $\alpha+\beta$ bits of information;
\item $\alpha\ge \KS(x\cnd y)$, since the bottom sender and the receiver together initially know only $y$, and should reconstruct $x$ getting $\alpha$ bits of information;
\item $\beta\ge\KS(y\cnd x)$, for symmetric reasons.
\end{itemize}

Slepian and Wolf~\cite{Slepian--Wolf} proved in the Shannon information theory setting that these information flow conditions are not only necessary, but also sufficient.  Algorithmic information versions of this theorem with increasingly better parameters  were obtained  in \cite{rom:j:slepwolf,rom:t:slepwolf, zim:c:kolmslepianwolf,bau-zim:t:univcompression}. The strongest result is in~\cite{bau-zim:t:univcompression} and it  says that the senders can use probabilistic polynomial-time algorithms to find messages of lengths $\alpha$ and $\beta$ (plus $\log^2 (n/\epsilon)$ overhead) that solve the Slepian--Wolf transmission problem with probability $(1-\epsilon)$, provided of course that $\alpha$ and $\beta$ satisfy the necessary information flow conditions.\footnote{Note that the senders do not  know each other data and, thus, the problem in~\cref{f:slepian-wolf} is that of \emph{distributed compression}. The interpretation of the Slepian-Wolf theorem  is  that, amazingly enough, distributed compression can achieve the same compression lengths as centralized compression (in which the senders share data and collaborate).}

We see that there are non-trivial results of both types (when the information flow conditions are sufficient and when they are not). So we come to  the following question:

\openq{
For which networks are the information flow conditions sufficient for the existence of a solution?
}

The relation between Shannon and algorithmic information theory approaches to network flow is not well understood. The Shannon information theory considers random variables as sources (and wants to reconstruct them at the destinations). To compare it with algorithmic setting, we consider strings obtained by $N$ independent samples of all the variables. If the information flow is achievable in the Shannon's sense, then one can use the same coding/decoding functions to solve the problem in the algorithmic setting (though maybe it does not give us logarithmic precision). The reverse direction looks even less clear: the reason is that the algorithmic setting is non-uniform and ``advices'' of logarithmic size are allowed in the intermediate vertices.

\openq{
Is the reverse statement true? If the information flow is not achievable in the Shannon's sense for independent samples of random variables, may it happen that with high probability the corresponding algorithmic information flow problem is solvable for those samples? More generally, what are the connections between algorithmic and Shannon's multi-source information theory?}

Even for the Slepian--Wolf network where there are results for both Shannon and algorithmic settings, the parallelism between them is quite perplexing: none of these theorems can currently be derived from the other one. 

\openq{
Is there a natural statement that would imply both the Slepian--Wolf theorem and its algorithmic version?
}

\section{Search-to-profile reductions}\label{s:srp2}

We started from inequalities for Kolmogorov complexities of tuples; they are universal statements saying that something is true for all strings. Then we considered some results that have  the more complicated $\forall A\, \exists B$-form: for example, the Slepian--Wolf theorem guarantees that for every strings $x,y$ and for all triples of numbers that satisfy the information flow constraints, one can find suitable messages for transmission. The obvious statement about the divisibility of information that we use as the opening example in Section~\ref{sec:info-theoretic} also has this form.

We can associate to a true $\forall A\, \exists B$-statement the natural \emph{search} problem: on input $A$, find a $B$ that certifies the truth of the statement, and, furthermore,  one may ask whether this problem can be effectivized, i.e.,  
(a) whether  an object $B$  can be found effectively, or even better, 
(b) whether it can be found efficiently. 
Since the complexity function is non-computable, it is natural to assume that the complexity profile of the input strings (from $A$) is also given. For some search problems the answer to question (a) is now positive in a rather trivial way. This is so, because for some problems, if we have the profile, we can effectively enumerate a list of ``suspects" that   is guaranteed to contain an object $B$ and, therefore, eventually we will find $B$. This type of  search is, alas, very slow.  For other problems, even with the profile in our hands, we cannot effectively enumerate a list of suspects as above and, for them, even a slow search procedure (not to mention an efficient one) would be remarkable.  Thus the real question is: For a $\forall A\, \exists B$-problem, if we are given the profile of the objects in $A$,  is there a non-trivial  search algorithm that finds  $B$?  In case the answer to this question is `yes', we say that the problem admits a \emph{\srp}  reduction.

For illustration, suppose we want to find a shortest or close to shortest program  of a string $x$ (arguably the primary search problem in Kolmogorov complexity). As it is well-known, there  is no algorithm for this task  (and this remains true even if we weaken a lot the optimality requirement, see for example~\cite[Lemma 6.2]{bau-zim:t:linlist}). On the other hand, having the profile of the input (which in this case consists of only $\KS(x)$) helps because we can run in parallel all programs of length $\KS(x)$ until one of them produces~$x$. Running in parallel all programs of the given length is an extremely long procedure.\footnote{This is an effective procedure but its runtime is larger than any computable function. An effective procedure with a non-computable runtime might look contradictory. But there is no paradox since the procedure  works with the promise that its input is $(x, \KS(x)$).}  Surprisingly, if we are content to find a program for $x$ whose length may be  slightly greater than the optimal value (with an overhead of size $O(\log^2 |x|)$) and also allow a small probability of failure,  then we can do the search in \emph{probabilistic polynomial time}.

This efficient procedure actually works  for the following  more general problem: Find a close to shortest program for a string $x$ conditioned by a string $y$ (the optimal size of such a program is by definition $\KS(x \cnd y)$). Muchnik's Theorem, discussed on page~\pageref{p:docexchange}, says that there exists an algorithm which on input $x$, and provided with some help information about the pair $(x, y)$ of length  only $O(\log \KS(x \cnd y))$ bits, returns a program of $x$ conditioned by $y$ of length  $\KS(x \cnd y)+O(\log \KS(x \cnd y))$. The help information in the original version of Muchnik's Theorem is a mystery, as it arises from a technical component of the proof given in~\cite{muchnik-conditional}.\footnote{In the papers~\cite{bfl:j:boundedkolmogorov,mus-rom-she:j:muchnik, bmvz:j:shortlist,teu:j:shortlists,zim:c:shortlistshortproof}, which (roughly speaking) look at this problem from various angles, the help information is the seed of an extractor, or of  a disperser (these are functions studied in the theory of pseudorandomness).}  Arguably, a more natural version of this result was proven in~\cite{bau-zim:c:linlist, bau-zim:t:univcompression}: there exists a probabilistic polynomial time algorithm that, on input $x$ and an integer $k$ with $k \ge \KS(x \cnd y)$,  outputs with probability $1-\varepsilon$ a program $p$ for $x$ given $y$ of length $|p| \le k + O(\log^2 ( |x|/\varepsilon))$. This is relevant for the ``document exchange" problem from page~\pageref{p:docexchange} in which Alice wants to send a document $x$ to Bob, who has an older version $y$. Even though Alice does not know $y$, she may know $\KS(x \cnd y)$ or an upper bound $k$ for it, i.e., she may have an idea about how ``far" is $x$ from $y$.\footnote{Still, unfortunately, the \emph{decoding} process, i.e., running the program $p$, is not efficient. This is unavoidable, since there are strings for which the bounded-time complexity is significantly higher than the unbounded one.} Note that in the above result  the ``help information" $\KS(x \cnd y)$ can be determined (with logarithmic precision)  from the profile of the pair $(x,y)$. Thus, the problem of finding a short program for $x$ given $y$, admits a \emph{\srp} reduction. 

In a similar way, the results on the Slepian-Wolf coding problem from~\cite{zim:c:kolmslepianwolf,bau-zim:t:univcompression} mentioned in~\cref{s:comm},  imply that this problem admits a \emph{\srp} reduction. In general, it may happen that all the results of the considered type can be effectivized. One may formulate the following bold conjecture: 

\begin{quote}
\emph{Search-to-profile reduction}. Given a tuple of strings $(x_1,\ldots,x_k)$ and the profile of a bigger tuple $(x_1,\ldots,x_k,y_1,\ldots,y_m)$ for some $y_1,\ldots,y_m$ (but not the $(y_1,\ldots,y_m)$ themselves), one can find (via a polynomial probabilistic algorithm) some $y_1',\ldots,y_m'$ such that the tuple $(x_1,\ldots,x_k,y_1',\ldots,y_m')$ has profile logarithmically close to the one that is given.
\end{quote}

\openq{Is this conjecture true or false? If it is false (it seems too optimistic to be true), what weaker versions hold true? }

There are several special cases when this question could be asked. For example, one may ask whether one can find effectively the messages for the Gray--Wyner information transmission problem when the pair $(x,y)$, its profile, and a point in its  Gray--Wyner profile are given. Similar questions may be asked for Wyner profile and extended profile; see also \ref{ak} in Section~\ref{s:profile-reduction}.
Note that for the Gray--Wyner profile the messages could be found by exhaustive search since the requirements are computably enumerable. One can also restrict in the Gray--Wyner problem (\cref{f:graywyner}) the sum of the lengths of all three messages $p,q,r$ and the length of the message $q$ that both recipients receive; this also leads to computably enumerable requirements (and therefore exhaustive search is possible). Similarly, a string $z'$ in \ref{ak} (if it exists) can be found  by  brute force search given $x$ and $y$ and the complexity profile of $(x,y,z)$. However, even if such a $z'$ exists, it is not clear whether it can be found efficiently.

In the previous questions we assumed that we are only given the complexity profile for \emph{one} specific tuple of strings. We can extend the setting and consider an oracle that computes the complexity function $\KS(x)$ for every string $x$.  It is known, for example, that  this oracle permits to find in polynomial time, for each $n$, an incompressible $n$-bit string $x$, see~\cite{bfnv}. Can we do more with this oracle? 
 
\openq{\label{qsrp-general-oracle}
 (Rather informal.)  Which search problem can be solved efficiently (say, in probabilistic polynomial time)  with  an oracle that computes $\KS$?  
}

We finish this section with the remark  that a similar reduction property  has been very recently noticed also in time-bounded Kolmogorov complexity~\cite{ila:c:minformula, lu-oli:c:coding}.

\section{Secret key agreement} \label{s:ska}

In information-theoretical cryptography,  the basic objects (messages, secret keys, shares of the secret, and so on) are represented by random variables, and the cryptographic strength of a secret key is measured by its Shannon entropy. This approach has many  practical and theoretical advantages, but  also  some limitations. For example, if a random number generator produces a secret key that consists of only zeros, we cannot use this specific key in practice. However, in the framework of the Shannon's theory we have no vocabulary to complain about such a key: an individual value of a random variable is irrelevant, only the probability distribution in the whole makes sense.  

In contrast, Kolmogorov complexity does allow us to talk about the ``quality'' of an individual instance of a secret key.  Surprisingly,  we still have no consistently developed framework of theoretical cryptography based on Kolmogorov complexity, even though some steps  were done by Antunes et al. in \cite{antunes}, who translated several cryptographic primitives (the one-time pad, secret sharing) in the language  of Kolmogorov complexity.   Some other results of this type were obtained by Muchnik in~\cite{muchnik-cryptography}.
 
We discuss here one specific cryptographic primitive, namely \emph{secret key agreement} (which is pretty well studied in the context of Shannon's approach (see the classical papers \cite{AhlswedeCsiszar1993,Mauerer1993} and more recent references in the surveys \cite{narayan-tyagi, wyner-in-tcs-syrvey}), and  recently in the framework of Kolmogorov complexity~\cite{romashchenko-zimand}). We focus on secret key agreement for two parties (Alice and Bob).
Let us assume that Alice and Bob possess some strings $a$ and $b$ that may be dependent (have mutual information). They want to use this fact to establish a common secret by communicating with each other via a public channel. They want to ensure that Eve, who initially has no information on $a$ and $b$ but overhears all communication messages, does not know anything about this common secret. In some natural setting it can be proven (under some minimal assumptions) that the size of this common secret can not exceed $I(a\cln b)$. 

To make the question more formal, we must explain  the ``communication'' between Alice and Bob in terms of Kolmogorov complexity. One way to do it is to say that every message (sent by Alice or Bob) must be ``simple'' (i.e., must have small Kolmogorov complexity) given the input data of this party and the previously received messages. More specifically, we consider a sequence of $O(1)$ strings $\Pi=(\pi_1,\ldots,\pi_k)$. Informally, $\pi_1$ stays for the first message of Alice, $\pi_2$ is the response of Bob, $\pi_3$ is the second message of Alice, etc.  Formally we require that $\pi_1$ is simple given $a$ (Alice has no other information when $\pi_1$ is sent), $\pi_2$ is simple given $b$ and $\pi_1$, then $\pi_3$ is simple given $a$ and $\pi_2$ (we may add also $\pi_1$ to the conditions, it does not change anything since $\pi_1$ is simple given $a$), etc. Then, if some string $c$ (common secret) is now known both to Alice and Bob (i.e., is simple with respect to $a,\Pi$ and with respect to $b,\Pi$), but the external observer has no information about it ($I(c\cln\Pi)\approx 0$), then the complexity of $c$ is at most $I(a\colon b)$ with logarithmic precision.

An alternative formalism uses the standard notion of a communication protocol. We assume Alice and Bob fix in advance a communication protocol (which can be randomized, with private or public sources of randomness), and then apply this protocol to the given inputs $a$ and $b$.   The transcript of the protocol $\Pi=(\pi_1,\ldots,\pi_k)$ is a sequence of messages computed according to the protocol rules. Now the number of messages (rounds) $k$ can be an arbitrary number, not necessarily a constant. We assume that Eve knows the protocol and can intercept each message (but, of course, Eve does not know $a$ and $b$). It turns out that in this setting the upper bound for the common secret remains the same: Alice and Bob cannot agree (with non-negligible probability) on a secret key of size larger than $I(a\colon b)$. And this size of the key is achievable: there is a communication protocol that permits to Alice and Bob to get a secret key of size approximately (again, with a logarithmic precision) $I(a\colon b)$. (Technically, we must assume that the parties  know in addition to $a$ and $b$ the complexity profile of $(a,b)$, which is only a logarithmic amount of information). The proofs of the lower and upper bounds can be found in \cite{romashchenko-zimand}. It is also known that the optimal communication complexity of this problem (in the worst case) is $\min\{ \KS(a \cnd b), \KS(b \cnd a)\}$, see \cite{gurpinar-romashchenko}.

The following questions remain unsolved.

\openq{ Find the optimal communication complexity for various important examples of $(a,b)$, e.g., for a pair of strings with a given Hamming distance.}

\openq{We said above ``of course, Eve does not know $a$ and $b."$ But what if Eve knows something, i.e., she has $c$ with $I((a,b) \cln c) > 0$? In other words, what is the optimal size of the secret key that can be achieved if Eve has some \emph{a priori} information about $a$ and $b$.}

For example, the following setting is motivated by quantum cryptography: Alice and Bob hold $n$-bit strings $x=x_1\ldots x_n$ and $y=y_1\ldots y_n$, Eve could somehow learn $x_i$ and $y_i$ for a small fraction of indices $i$, and in the remaining part (that Eve does not know) the bits $x_i$ and $y_i$ are significantly correlated. What can Alice and Bob (who do not know which bits are visible to Eve) do to create a common secret unknown to Eve, and what are the restrictions on the size of this secret and amount of communications needed?   

In the multi-party version of secret key agreement we know how to compute the optimal size of the common secret key that can be obtained by the parties, but, unlike the two-party version, we do not know the communication complexity.

\openq{Find the optimal communication complexity in the multi-party of secret key agreement (see \cite{romashchenko-zimand, gurpinar-romashchenko} for details). 
}

\section{Kolmogorov complexity in bipartite graphs} \label{s:binrel}

Previously we considered purely information-theoretic properties of finite objects without any structure on them. In this section we add such a structure in the form of a certain finite bipartite graph. This brings up the following generic question: given a bipartite graph and given a left vertex of some known complexity, what can we say about the distribution of complexities among its right neighbors? Let us look at several instances of this question.

Consider the graph where left and right vertices are $n$-bit strings, and $x$ and $y$ are connected by an edge if the Hamming distance $d_H(x,y)$ does not exceed some threshold $d$. Then the neighbors of some string $x$ form a Hamming ball of radius $d$ centered at $x$.  Consider the \emph{minimal} complexity of ($d$-) neighbors of $x$. For every $x$ this minimal value is a decreasing function of $d$. These functions may be different for different strings $x$; what functions may appear in this way? The answer is provided by algorithmic statistics  (see~\cite{algorithmic-statistics-survey} for details).

\openq{Consider now the \emph{maximal} complexity of a $d$-neighbor as a function of $d$, for a given vertex $x$. Which functions appear in this way?}

Some partial results are known. It was noted in~\cite{bfnv} that Harper's theorem (saying that Hamming balls have minimal neighborhood in the Boolean cube among all sets of the same size) gives a lower bound for this function (knowing the complexity of $x$, we can prove a lower bound for the maximal complexity in the $d$-neighborhood), and this bound is tight. The tight upper bound is also known --- as well as some other partial results recently obtained by Kozachinskiy and Vereshchagin (private communication).

Instead of minimal or maximal complexity, we may consider the \emph{typical} complexity of a $d$-neighbor. It is known (see~\cite{posobin}) that this notion is well defined, i.e., for every $x$ most of its $d$-neighbors have (almost) the same complexity. This complexity is (for every $x$) a function of $d$.

\openq{Which functions of $d$ can appear in this way?}

Some tight lower bounds for these functions are found in~\cite{posobin}.

More generally, one may investigate the possible distributions of complexity among the neighbors of a given vertex.

Another interesting graph is formed by lines and points on the affine plane. Consider some finite field $\mathbb{F}$ and an affine plane over $\mathbb{F}$. The points on this plane are pairs $(x,y)$ where $x,y\in\mathbb{F}$; the lines are defined by equations $y=kx+b$ and $x=c$ (vertical lines). A bipartite graph is formed: left vertices are lines, right vertices are points, and edges connect incident line and point. If the field $\mathbb{F}$ has about $2^n$ elements, then a random left vertex (line) has complexity about $2n$; the same is true for a random right vertex (point). A random edge has complexity about $3n$ (since the incidence condition subtracts $n$ bits of complexity).

We already have seen this graph in~\cref{s:extractci} when discussing common information: a random edge of it, considered as a pair of finite objects (line and point), has non-extractable common information. We asked there what is the Gray--Wyner profile (or extended profile, or Wyner profile) of a random edge.

It is known that the Gray--Wyner profiles of all random edges are the same. Indeed, it is shown in~\cite{muchnik-upper-semilattice,ilyaraz} that for edge-transitive graph of low complexity this profile is determined by the following combinatorial property: what is the maximal number of edges in a combinatorial rectangle of size $2^k\times 2^l$. But we do not know whether the extended profiles of random edges are the same. 

The question that we discussed for the Hamming graph, makes sense also for lines and points graph:

\openq{What distribution of complexities may appear among the neighbors of a line of a given complexity?}

Some partial results are known here. One can prove that the maximal complexity of a point on a line of complexity $k$ is $\min(k+n,2n)+O(\log n)$. This can be considered as a finite counterpart for the result about effective Hausdorff dimensions of points on a line of given effective Hausdorff dimension~\cite{lutz-stull}, though the argument is much easier.  

Note also that answers to all these questions may depend on the choice of a finite field $\mathbb{F}$. One may also consider an continuous version of the problem when we consider lines and points over $\mathbb{R}$ but with some finite precision. This setting looks closer to the question about Hausdorff dimension mentioned above, but the lower bound for the maximal complexity of points on a line (that we mentioned) does not work anymore.

\section{Randomness deficiencies}\label{sec:deficiencies}

This section is an outlier in the sense that here we speak about infinite objects (binary sequences). Still the questions we mention are not well studied in the algorithmic randomness theory because of their quantitative nature: we do not just look whether a sequence is random or not, but are interested in its \emph{randomness deficiency} (that is finite for random sequences and infinite for non-random ones). Let us explain what it means.

One of the most remarkable achievements of algorithmic information theory is the definition (that goes back to Martin-L\"of) of randomness for individual objects (infinite binary sequences) and its characterization (due to Schnorr and Levin, see~\cite{suv} for the history for these developments) in terms of complexity. One (probably the most popular) version of this characterization goes as follows: a sequence $\omega$ is Martin-L\"of random with respect to the uniform Bernoulli measure if and only if $\KP(x)\ge |x|-c$ for some $c$ and all prefixes $x$ of $\omega$. Here $\KP(x)$ is the prefix complexity of $x$, and $|x|$ stands for the length of $x$.

Soon it was realized (by Levin and G\'acs, see~\cite{gacs-exact,gacs-survey}) that this result has a natural quantitative version. One can define the \emph{universal expectation-bounded randomness test} $\omega\mapsto\mathbf{t}(\omega)$ as a maximal (up to a constant factor) lower semicomputable non-negative real-valued function on infinite sequences that has finite expected value (over uniform Bernoulli measure).\footnote{Lower semicomputability means that a function can be computably approximated from below; Bernoulli measure $B_p$ is the distribution of independent coin tossings with the probability of success $p$; when $p=1/2$, we speak about uniform Bernoulli distribution. See~\cite{gacs-survey} for details.}  The following result connects this test (with maximal possible $O(1)$-precision) with prefix complexity of prefixes of $\omega$:
\[
\log \mathbf{t}(\omega) = \max_{\text{$x$ is a prefix of $\omega$}} [|x|-\KP(x)].
\]
In exponential scale this formula can be rewritten as
\[
\mathbf{t}(\omega) =\max_{\text{$x$ is a prefix of $\omega$}}\frac{\mathbf{m}(x)}{P(x)}
\]
Here $\mathbf{m}(x)$ is the so-called \emph{discrete a priori probability}, the maximal lower semicomputable non-negative real function on strings that has finite sum; as Levin and later Chaitin have noted, $\mathbf{m}(x)=2^{-\KP(x)}$ with maximal possible precision (up to $O(1)$-factors in both direction). By $P(x)$ we denote the uniform Bernoulli measure of the set of all extensions of the string $x$, i.e., $P(x)=2^{-|x|}$. In this form this formula allows generalization for the case of arbitrary computable measure $P$; one can also replace maximum by sum:
\[
\mathbf{t}(\omega) =\sum_{\text{$x$ is a prefix of $\omega$}}\frac{\mathbf{m}(x)}{P(x)}
\]
This remarkable (in its precision) result still leaves many question open.
\openq{The original Martin-L\"of definition corresponds to a slightly different notion of randomness test: a \emph{universal probability-bounded test} that is a maximal non-negative lower semicomputable function $\mathbf{t}$ such that the event $\mathbf{t}(\omega)\ge c$ has probability at most $1/c$ for every positive $c$. Is it possible to find a similar characterization of this kind of randomness deficiency in terms of complexities (plain, prefix, monotone or some other ones)? It is known that expectation-bounded deficiency (in logarithmic scale, defined as a logarithm of the corresponding universal tests) and probability-bounded deficiencies are close to each other, but they do not coincide~\cite{novikov} and it is not known whether one can be somehow expressed in terms of the other. We do not know also whether it is possible to find two sequences $\omega_1$ and $\omega_2$ such that the expectation-bounded deficiency of $\omega_1$ is bigger than for $\omega_2$, while the probability-bounded deficiency of $\omega_2$ is bigger than for~$\omega_1$.
}
\openq{There are other version of Schnorr--Levin characterization of random sequences: for example, they are sequences for whose prefixes $x$ the difference between $|x|$ and monotone complexity $\KM(x)$ is bounded. Here the monotone complexity can be understood in the sense of Levin; it can be also replaced by a priori complexity $\KA(x)$. Do these results have a quantitative version similar to the formula for randomness deficiency mentioned above?
}
\openq{One can go in the other direction: starting for a deficiency function $d(\omega)$ defined (in some way) on infinite sequences $\omega$, one can define deficiency of a finite sequence $x$ as the minimal deficiency of all its infinite extensions. This idea goes back to Kolmogorov: seeing a finite sequence of heads and tails, we reject it as a possible outcome of a fair coin if all its possible extensions look dubious as infinite random sequences. In this way we get a lower semicomputable deficiency function on finite strings. Can we give a characterization of this function (with logarithmic or better precision) in terms of complexities of finite strings (using some version of complexity)?
}
\openq{The previous question can be extended for randomness with respect to \emph{classes} of measures. Consider an effectively closed class of measures, for example, the class of Bernoulli measures $B_p$ for all $p\in[0,1]$ or the class of all shift-invariant measures on infinite sequences. Then one can naturally define universal tests; for example, the universal expectation-bounded test for Bernoulli measures is the maximal lower semicomputable non-negative real function on infinite binary sequences whose expectation according to every Bernoulli measure is at most~$1$ (see~\cite{gacs-survey} for the exact definition and discussion). Can we relate those tests to complexities of finite strings? A natural idea (that goes back to Kolmogorov) is to compare $\KS(x)$ for a finite string $x$ to the logarithm of the number of all strings of the same length (as $x$) that have the same numbers of zeros and ones. In general, is it possible to decide whether a bit sequence $\omega$ is Bernoulli random (has finite values of the universal Bernoulli test) or not, knowing only the complexities of all its prefixes and the number of heads and tails in all its prefixes?

For the class of all shift-invariant measures the situation is much more unclear: what properties of a finite sequence of heads and tails make us reject the conjecture that it is generated by some shift-invariant random process?
}
\openq{One can try to start with Levin--G\'acs formula for randomness deficiency and go in a different direction. This formula is a quantitative version of a Schnorr--Levin ``qualitative'' complexity characterization of randomness. There are some other results about randomness that have natural quantitative versions (e.g., the van Lambalgen theorem about randomness of pairs~\cite{bauwens-lambalgen}; the image randomness characterization~\cite{bienvenu-hoyrup}). Can one develop this theory systematically to get quantitative versions of other qualitative results?}

\section*{Acknowledgments} We are grateful to Bill Gasarch for inviting us to write this survey, and to Lance Fortnow for his useful suggestions.

\end{document}